\documentclass[conference,letterpaper]{IEEEtran}
\usepackage{cite}
\usepackage{amsmath,amssymb,amsfonts}
\usepackage{algorithmic}
\usepackage{algorithm}
\usepackage{graphicx}
\usepackage{textcomp}
\usepackage{xcolor}
\usepackage{cuted}

\begin{document}

\title{Decision-Focused Learning for Cyber-Physical Risk-Aware Power Grid Operations}

\author{
\IEEEauthorblockN{Afzal Hossain Sohag}
\IEEEauthorblockA{
\textit{Department of Computer Science}\\
\textit{Florida Polytechnic University}\\
Lakeland, FL, USA\\
asohag3061@floridapoly.edu
}
\and
\IEEEauthorblockN{Ayesha S. Dina}
\IEEEauthorblockA{
\textit{Department of Computer Science}\\
\textit{Florida Polytechnic University}\\
Lakeland, FL, USA\\
adina@floridapoly.edu
}
\and
\IEEEauthorblockN{Ebisa D. Wollega}
\IEEEauthorblockA{
\textit{Department of Industrial Engineering}\\
\textit{Florida Polytechnic University}\\
Lakeland, FL, USA\\
ewollega@floridapoly.edu
}
}

\maketitle

\begin{abstract}
Power grid cyber-physical security requires intrusion detection methods that
account for the operational consequences of detection errors. However, the existing
intrusion detection approaches primarily focus on the predictive performance of learning algorithms, while downstream operational costs are often optimized independently, which can lead to inferior operational performance. To help alleviate this problem, we propose a novel optimization-learning cyber-physical-risk-informed framework that communicates iteratively to minimize power grid operational costs. This framework translates machine learning prediction errors into power grid operational cyber-physical-risk costs. Experimental results on two power grid operation datasets, Industrial Control System (ICS) Cyber Attack and Sherlock, show that the proposed approach improves power grid operation costs compared with the traditional approach. 
\end{abstract}

\begin{IEEEkeywords}
Intrusion Detection, Decision-Focused Learning, Cyber-Physical-Risk, Power Grid Operations.
\end{IEEEkeywords}
\footnotesize
This work has been submitted to the IEEE for possible publication.
Copyright may be transferred without notice, after which this version
may no longer be accessible.
\normalsize

\section{Introduction}

Modern power systems are evolving into highly interconnected smart grids that integrate communication, automation, distributed energy resources, and real-time monitoring. The increased deployment of sensors, smart meters, supervisory control and data acquisition (SCADA) systems, and intelligent control devices has significantly improved operational efficiency, reliability, and energy management capabilities. However, this digitalization also introduces security challenges because power grid operation increasingly depends on communication networks and cyber-physical interactions \cite{sun2018cyber,krause2021cybersecurity}.

Unlike conventional isolated power systems, power grids continuously exchange operational information among generators, substations, control centers, and consumers. This interconnection increases exposure to cyber-physical threats such as false data injection (FDI), denial-of-service (DoS), malware, and communication tampering \cite{liu2011false,yan2012survey}. Existing studies have shown that attackers can manipulate sensor measurements or communication signals to disrupt state estimation, destabilize system operation, and interfere with economic dispatch procedures \cite{tan2015integrity,yang2020distributed,liu2011false}. Such intrusions can cause poor operational decisions, financial losses, cascading failures, and service interruptions. Among these threats, FDI attacks are a major cybersecurity risk in modern smart grids \cite{liu2011false,yan2012survey}. Because operational decisions rely on accurate state estimation and real-time measurements, compromised data can directly affect generator scheduling, load balancing, and grid optimization.

To mitigate these risks, several cyber-physical security solutions have been proposed, including intrusion detection systems (IDSs), anomaly detection, encryption, and secure communication architectures \cite{sun2018cyber,krause2021cybersecurity,yan2012survey}. Machine learning-based intrusion detection approaches can identify complex attack patterns and abnormal behaviors in cyber-physical environments, including suspicious measurements in substation
communication systems \cite{muller2022assessment,sahani2023machine,haque2020machine,sourav2023machine}. Existing studies have investigated supervised learning, unsupervised learning, and process-aware intrusion detection methods for detecting cyber-physical attacks in power grid systems \cite{muller2022assessment,sahani2023machine,wagner2024sherlock}. Although these approaches have demonstrated promising detection performance, many existing methods primarily optimize predictive metrics such as accuracy, recall, precision, and F1-score without considering the downstream operational cost of prediction errors on power system decision-making.

Significant research has also been conducted on resilient economic dispatch, robust optimization, and secure operational planning under attacks and uncertain operating conditions \cite{yang2020distributed,sun2017robust}. Integrity and FDI attacks can destabilize dispatch decisions and disrupt electricity market operation \cite{liu2011false,tan2015integrity,yang2020distributed}. Robust optimization approaches have been proposed to improve reliability and robustness in power systems \cite{sun2017robust}. However, existing studies have investigated cyber-physical attack detection and resilient economic dispatch, with comparatively less emphasis on how detection errors translate into operational cost \cite{li2025resilient, huang2020real,wlazlo2021man}.

Decision-focused learning framework integrates machine learning with downstream optimization objectives \cite{elmachtoub2022smart,donti2017task,bertsimas2020predictive,mandi2024decision}. Unlike traditional predict-then-optimize pipelines, these approaches directly optimize decision quality rather than prediction accuracy alone. Although decision-focused learning has shown promise in operational optimization, limited research has integrated cyber-physical-risk-aware intrusion detection, and power grid operations optimization.

To address these limitations, this paper proposes a cyber-physical-risk-aware intrusion detection and power grid optimization framework for cyber-physical power systems. The proposed framework integrates machine learning-based intrusion detection with power grid operational optimization by transforming intrusion detection misclassification outcomes into generator-level cyber-physical-risk costs that guide power grid operational planning. Unlike traditional intrusion detection systems that primarily evaluate predictive performance, the proposed framework incorporates asymmetric misclassification penalties to represent the different operational consequences of false positives and false negatives within the power grid optimization.

The framework is evaluated on two benchmark cyber-physical power-system datasets to examine the relationship between intrusion detection performance and operational cost.
The main contributions of this work are summarized as follows:

\begin{itemize}

\item Development of an optimization-informed cyber-physical-risk framework that integrates machine learning-based intrusion detection with power grid operational optimization.

\item Transformation of intrusion detection misclassification outcomes into generator-level cyber-physical-risk costs for power grid operational optimization.

\item Development of an iterative feature-reweighting mechanism that uses learned IDS parameters to update the feature representation for subsequent IDS retraining.

\item An evaluation of intrusion detection performance and operational cost on two benchmark cyber-physical power-system datasets with different grid topologies and attack scenarios.

\end{itemize}

\section{Problem Formulation}

\subsection{Power System Model}

The proposed framework considers a cyber-physical power system in which operational decisions are jointly influenced by conventional generation costs and cyber-physical security risk. Building on prior work on cyberattack-resilient economic dispatch and power-system optimization \cite{yang2020distributed, sun2017robust, huang2020real}, the framework integrates machine learning-based intrusion detection with operational optimization to incorporate cyber-physical-risk information into dispatch decisions.

\begin{figure}[t]
\centering
\includegraphics[width=\columnwidth]{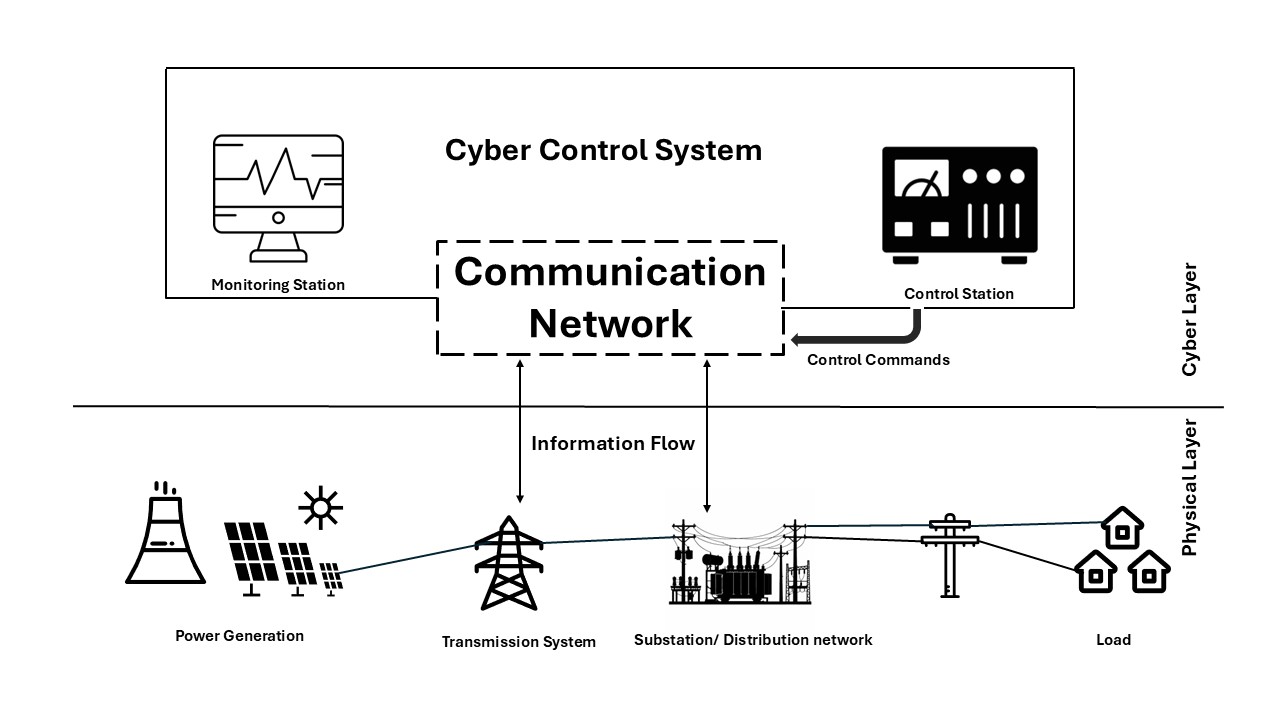}
\caption{Cyber-physical power system architecture illustrating the cyber-physical interaction underlying the research problem.}
\label{fig:framework}
\end{figure}

Fig.~\ref{fig:framework} illustrates the cyber-physical interaction between the communication and control infrastructure and the physical power system. Operational measurements are transmitted from the physical layer through the communication network to the control system, while control commands are returned to the physical layer. Cyber-physical attacks targeting this information flow can compromise operational measurements and potentially affect subsequent power grid decisions.

Let \(G\) denote the set of generators and \(L\) denote the set of system loads. The power allocation decision variable is defined as

\begin{equation}
x_{ij} \geq 0, \quad \forall i \in G,\; j \in L
\label{eq:nonnegative}
\end{equation}

where \(x_{ij}\) represents the amount of power supplied from generator \(i\) to load \(j\).

Each generator is associated with an operating cost  \(C_i\), and each generator-to-load line has an associated transmission cost \(T_{ij}\). During the iterative framework, generator operating costs are adjusted according to the cyber-physical risk estimated from intrusion detection outcomes. The resulting cyber-physical-risk-adjusted generator costs are subsequently incorporated into the power grid operational optimization model presented in Section~\ref{sec:optimization}.

\subsection{Misclassification Cost Formulation}

The proposed framework incorporates intrusion detection outcomes into the
power-grid operational optimization process by assigning operational
penalties to different types of prediction errors. False-positive (FP) and
false-negative(FN) outcomes can have different consequences in cyber-physical
power systems, particularly because undetected compromises may propagate
into subsequent operational decisions \cite{ding2020cyber,elkan2001foundations}.

Let \(y_k\) denote the true label of sample \(k\), and let \(\hat{y}_k\)
denote the predicted label generated by the intrusion detection model.
Instead of treating all prediction errors equally, different penalties are
assigned to the possible misclassification outcomes:

\begin{equation}
c_k =
\begin{cases}
C_{FN}, & y_k = 1 \text{ and } \hat{y}_k = 0 \\
C_{FP}, & y_k = 0 \text{ and } \hat{y}_k = 1 \\
0, & \text{otherwise}
\end{cases}
\end{equation}

where \(C_{FN}\) and \(C_{FP}\) represent the false-negative and
false-positive penalties, respectively. The framework allows asymmetric
misclassification costs, with configurations assigning a larger penalty
to false negatives to represent the potentially greater operational
consequences of undetected attacks.

\subsection{Generator-Level Cyber-physical-Risk Aggregation}

Each sample \(k\) is assigned to a generator \(g(k)\in G\) based on the scenario-to-generator mapping defined during dataset preprocessing. This mapping enables intrusion detection outcomes to be linked to individual generators and allows sample-level misclassification penalties to be aggregated into generator-level misclassification costs.

The total misclassification cost associated with generator \(i\) is computed as

\begin{equation}
M_i =
\sum_{k:g(k)=i} c_k
\end{equation}

where \(M_i\) represents the total misclassification cost associated with generator \(i\).

To obtain comparable cyber-physical-risk contributions across generators, the aggregated costs are normalized as

\begin{equation}
\tilde{M}_i =
\frac{M_i}
{\sum_{i\in G}M_i}
\end{equation}

where \(\tilde{M}_i\) denotes the normalized cyber-physical-risk contribution associated with generator \(i\). The normalized values are subsequently used to compute the cyber-risk-adjusted operating costs presented in the following subsection.

\subsection{Cyber-Risk Cost Integration}

The normalized cyber-physical-risk associated with each generator is transformed into a cyber-physical-risk cost that can be incorporated into the power grid operational optimization model:

\begin{equation}
C_i^{\mathrm{risk},t}
=
\alpha \cdot \tilde{M}_i
\end{equation}

where $\alpha$ is a scaling parameter controlling the influence of cyber-physical risk on power grid operational optimization.  $C_i^{\mathrm{risk},t}$ represents the generator-level cyber-physical-risk cost computed during iteration $t$ of the learning step. 

Before initiating the iterative learning process, the generator-level cyber-physical-risk costs are computed using the baseline intrusion detection model. For each hyperparameter configuration, a baseline Logistic
Regression model is first trained to obtain the baseline F1-score and
model coefficients, which are then used to
identify false-positive and false-negative samples and compute the
corresponding generator-level cyber-physical-risk costs. The corresponding baseline generator operating cost is defined as

\begin{equation}
C_i^{\mathrm{baseline}}
=
C_i
+
C_i^{\mathrm{risk},0}
\end{equation}

where $C_i^{\mathrm{risk},0}$ denotes the baseline generator-level cyber-physical-risk cost. This baseline cost serves as the reference for evaluating operational-cost changes during the iterative framework. Using these baseline effective generator costs in the dispatch optimization model, the resulting objective value is denoted by $Z^{\mathrm{baseline}}$.

During each iteration, the intrusion detection model is retrained using feature weights derived from the logistic regression coefficients of the preceding iteration, producing updated predictions and generator-level cyber-physical-risk estimates. Consequently, the generator operating cost is updated as

\begin{equation}
\label{eq:newcost}
C_i^{\mathrm{new}}
=
C_i
+
C_i^{\mathrm{risk},t}
\end{equation}

where $C_i$ denotes the original operating cost of generator $i$. The updated generator costs are subsequently incorporated into the power grid operational optimization model, increasing the effective operating cost of generators with higher cyber-physical-risk penalties.

\subsection{Cyber-Risk-Aware Dispatch Optimization}
\label{sec:optimization}

The proposed framework integrates cyber-physical-risk information into the dispatch optimization model by augmenting the generator cost with cyber-physical-risk penalties derived from intrusion detection outcomes. The objective is to minimize the total operational cost while accounting for potential cyber-physical-security risks associated with individual generators.

The cyber-physical-risk-adjusted generator cost \(C_i^{new}\), defined in \eqref{eq:newcost}, is incorporated into the dispatch optimization problem. Using the adjusted generator cost, the dispatch optimization problem is formulated as:

\begin{equation}
Z = \min \sum_{i \in G}\sum_{j \in L}
\left(C_i^{new}+T_{ij}\right)x_{ij}
\end{equation}

subject to the following constraints.

Generator capacity constraints ensure that the total power supplied by each generator does not exceed its maximum generation capability:

\begin{equation}
\sum_{j \in L} x_{ij}
\leq
GenCap_i,
\quad
\forall i \in G
\end{equation}

where \(GenCap_i\) denotes the capacity limit of generator \(i\).

Transmission capacity constraints limit the amount of power that can be transmitted through each generator-load connection:

\begin{equation}
x_{ij}
\leq
LineCap_{ij},
\quad
\forall i \in G, j \in L
\end{equation}

where \(LineCap_{ij}\) represents the transmission capacity between generator \(i\) and load \(j\).

Load demand constraints ensure that system demand is satisfied:

\begin{equation}
\sum_{i \in G} x_{ij}
\geq
D_j,
\quad
\forall j \in L
\end{equation}

where \(D_j\) denotes the demand requirement of load \(j\).

By incorporating cyber-physical-risk penalties into the objective function, the proposed framework assigns higher costs to generators associated with greater cyber-physical risk, which can influence their dispatch subject to the system constraints.

\subsection{Iterative Feature-Reweighting Learning Process}

The proposed framework iteratively updates feature weights using the logistic regression coefficients learned in the previous iteration.

Let $\boldsymbol{w}^{(t)}$ denote the feature-weight vector obtained from the
intrusion detection model at iteration $t$. The baseline IDS is first trained, producing the initial feature-weight vector
$\boldsymbol{w}^{(0)}$. At each subsequent iteration, the training feature
matrix is transformed using the feature weights obtained from the previous
iteration:

\begin{equation}
\mathbf{X}^{(t)} =
\mathbf{X} \odot \boldsymbol{w}^{(t-1)},
\end{equation}

where $\mathbf{X}$ denotes the original training feature matrix and $\odot$
represents element-wise multiplication. The IDS is then retrained using the
transformed feature matrix $\mathbf{X}^{(t)}$ and training labels $\mathbf{y}$:

\begin{equation}
\boldsymbol{w}^{(t)}
=
\operatorname{IDS}
\left(
\mathbf{X}^{(t)},\mathbf{y}
\right).
\end{equation}

In the proposed implementation, the IDS is realized using logistic regression,
and $\boldsymbol{w}^{(t)}$ corresponds to its learned coefficient vector.
For each iteration, the retrained IDS generates predictions
$\hat{y}_k^{(t)}$, which are used to determine false-positive and false-negative
outcomes. These outcomes are used to recompute generator-level cyber-physical-risk
costs, which are incorporated into the dispatch optimization.

\subsection{Framework Algorithm}

The overall workflow of the proposed framework is summarized in Algorithm~1.

% \vspace{0.06in}
\begin{algorithm}[!t]
\caption{Optimization-Learning Cyber-Physical-Risk-Informed Framework}
\label{alg:framework}
\begin{algorithmic}[1]

\STATE \textbf{Input:} Dataset $\mathcal{D}=(\mathbf{X},\mathbf{y})$,
generator assignments $g$, power grid parameters,
$\alpha$, $C_{FP}$, $C_{FN}$

\STATE Train baseline IDS and obtain coefficient vector
$\boldsymbol{w}^{(0)}$

\STATE Compute baseline cyber-physical-risk cost and obtain
$Z^{\mathrm{baseline}}$

\STATE Initialize feature weights
$\boldsymbol{w}\leftarrow \boldsymbol{w}^{(0)}$;
$Z^{\mathrm{best}} \leftarrow Z^{\mathrm{baseline}}$

\FOR{$t=1,\ldots,T$}

    \STATE Learn IDS using $\mathbf{X}^{(t)}$ and $\mathbf{y}$;
    obtain coefficient vector $\boldsymbol{w}^{(t)}$

    \STATE Estimate $\hat{y}_k^{(t)}$ and compute $c_k$

    \STATE Compute $M_i$, $\tilde{M}_i$, and
    $C_i^{\mathrm{risk},t}$

    \STATE Compute $C_i^{\mathrm{new}}$

    \STATE Obtain $Z^{(t)}$ and dispatch plan
    $x_{ij}^{(t)}$

    \STATE Update feature weights
    $\boldsymbol{w} \leftarrow \boldsymbol{w}^{(t)}$

    \STATE $Z^{\mathrm{best}} \leftarrow
    \min\{Z^{\mathrm{best}},Z^{(t)}\}$

\ENDFOR

\RETURN Power grid operational plan associated with
$Z^{\mathrm{best}}$

\end{algorithmic}
\end{algorithm}
The algorithm shows the iterative interaction between intrusion detection,
generator-level cyber-physical-risk estimation, and cyber-physical-risk-aware dispatch
optimization. The minimum operational cost across the predefined iterations
is retained as $Z^{\mathrm{best}}$.

\section{Experimental Setup}

\subsection{Dataset and Preprocessing}

We evaluate the proposed framework on the ICS Power System Dataset~\cite{7063234} and the Sherlock 01-Basic dataset~\cite{wagner2024sherlock}. The ICS dataset comprises 78,377 samples with 128 input features. Scenario identifiers are converted into binary benign/attack labels, and samples are assigned to two generator groups using the scenario-based generator mapping. Missing values are replaced using feature-wise median imputation, followed by standardization using statistics fitted on the training set.

The Sherlock 01-Basic dataset contains 43,204 samples with 470 state features representing grid measurements and states with binary attack labels. Missing values are imputed using training-set medians, followed by training-set standardization. For generator-level cyber-physical-risk aggregation, each sample is assigned to the generator with the maximum standardized active-power measurement. 
\begin{figure*}[t]
    \centering
    \includegraphics[width=\textwidth]{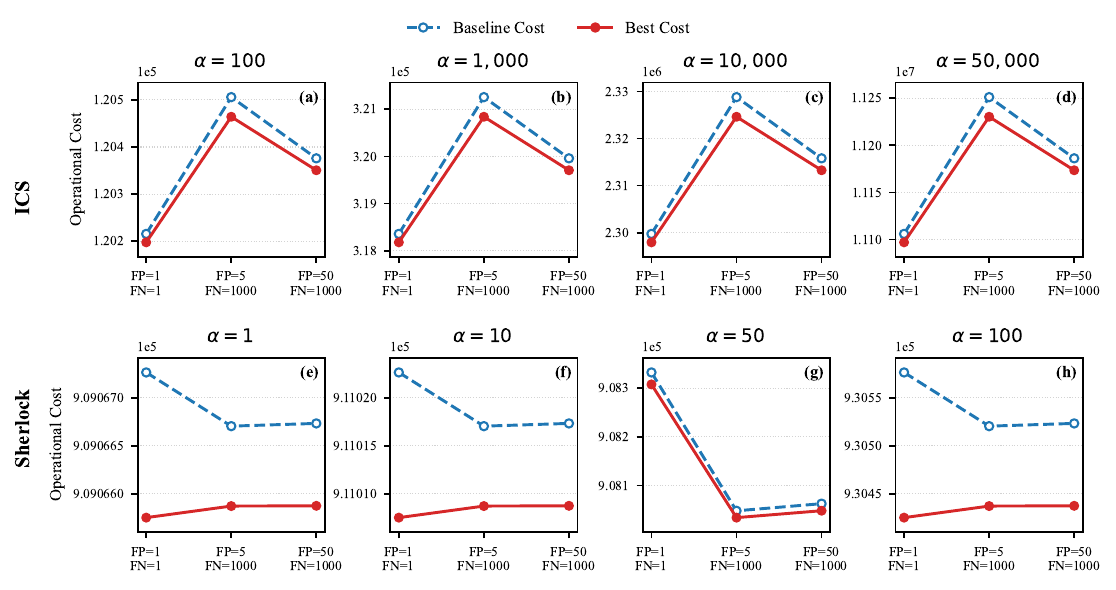}
    \caption{Sensitivity of operational cost to the cyber-physical-risk scaling
    parameter $\alpha$ for the ICS dataset (top row) and Sherlock dataset
    (bottom row). Dashed lines represent the baseline cost, while solid
    lines represent the minimum cost obtained during the iterative process.}
    \label{fig:alpha_cost_comparison}
\end{figure*}
\subsection{Experimental Configuration}

The intrusion detection component learns through logistic regression, while the cyber-physical-risk-aware economic dispatch is optimized using a linear programming formulation via the Gurobi optimizer \cite{gurobi}. Power generator cyber-physical-risk costs are added to the operating costs before dispatch. The dispatch model uses two generators for the ICS dataset, and seven assigned generators for the Sherlock dataset. A grid search was conducted over the predefined values of $\alpha$, $C_{FP}$, and $C_{FN}$ to evaluate their effects on operational cost.

Each parameter configuration is evaluated for 100 framework iterations. Dataset-specific generator capacities, operating costs, and load demands are configured according to the respective experimental models. The experimental parameters are summarized in Table~\ref{tab:parameters}.

\begin{table}[!t]
\caption{Experimental Parameters}
\label{tab:parameters}
\centering
\begin{tabular}{lc}
\hline
\textbf{Parameter} & \textbf{Value} \\
\hline
Baseline Learning Iterations & 2000\\
Optimization-Learning Framework Iterations & 100\\
False Negative Cost ($C_{FN}$) & Tuned\\
False Positive Cost ($C_{FP}$) & Tuned\\
Cyber-Risk Scaling ($\alpha$) & Tuned\\
Optimization Solver & Gurobi\\
\hline
\end{tabular}
\end{table}

\subsection{Cyber-Physical Risk Hyperparameter Selection}

Three hyperparameters were tuned: the cyber-physical-risk scaling factor ($\alpha$), the false-positive cost ($C_{FP}$), and the false-negative cost ($C_{FN}$). An exhaustive grid search was performed over predefined candidate values. Each parameter configuration was evaluated using the same 100-iteration framework, and the configuration yielding the minimum operational dispatch cost was selected as the final setting, which power grid operators can use for decision making.

\subsection{Evaluation Metrics}

The proposed framework was evaluated from both intrusion detection and power-system operation perspectives. Intrusion detection performance was measured using the F1-score, which provides a balanced assessment of precision and recall under class-imbalanced conditions.

Operational performance was evaluated using the total dispatch cost obtained from the cyber-physical-risk-aware optimization model, incorporating generation, transmission, and cyber-physical-risk-adjusted operating costs. Since operational cost is the primary objective of the framework, the configuration with the minimum operational dispatch cost was selected during hyperparameter tuning.
The model baseline performs intrusion detection and cyber-physical-risk cost computation economic dispatch without iterative feature reweighting. Its effectiveness was assessed by comparing the F1-score and operational dispatch cost to the proposed cyber-physical-risk-informed optimization-learning framework.
\begin{table}[t]
\caption{Cost Improvement Across FP/FN Configurations}
\label{tab:alpha_improvement}
\centering
\begin{tabular}{ccc}
\hline
\textbf{$\alpha$} &
\textbf{ICS (\%)} &
\textbf{Sherlock (\%)} \\
\hline
1       & 0.00046 & 0.00076 \\
10      & 0.00448 & 0.00754 \\
50      & 0.01803 & 0.01654 \\
100     & 0.03734 & 0.07376 \\
1,000   & 0.14001 & 0.00000 \\
5,000   & 0.18529 & 0.00000 \\
10,000  & 0.19309 & 0.00000 \\
50,000  & 0.19983 & 0.00000 \\
\hline
\end{tabular}
\end{table}

\section{Results and Discussion}

\subsection{Cyber-Physical-Risk Hyperparameter Sensitivity Analysis}

A sensitivity analysis was conducted to examine the effects of the
cyber-risk scaling parameter $\alpha$ and the false-positive and
false-negative misclassification costs on operational dispatch cost.
For each hyperparameter configuration, the minimum operational cost
observed over the 100 framework iterations was compared with its
corresponding baseline cost.

Figure~\ref{fig:alpha_cost_comparison} illustrates the sensitivity of
operational cost to $\alpha$ for the ICS and Sherlock datasets. The top
row presents the ICS results, while the bottom row presents the
Sherlock results. For each $\alpha$ value, the baseline and minimum
operational costs are compared for three representative FP/FN
configurations.
For the ICS dataset, cost improvement increases with $\alpha$, but the
additional gains become limited beyond $\alpha=5{,}000$. For the
Sherlock dataset, the maximum cost improvement increases up to
$\alpha=100$, while no improvement is observed for
$\alpha\geq1{,}000$. These results indicate that the effect of
cyber-risk scaling depends on the dataset and misclassification-cost
configuration.
The results show that the effect of $\alpha$ on operational cost varies
across datasets and FP/FN cost configurations.

\subsection{ICS Dataset Results}

Table~\ref{tab:alpha_improvement} summarizes the maximum cost
improvement obtained across the FP/FN configurations for each
$\alpha$. The improvement increased from 0.00046\% at
$\alpha=1$ to 0.19983\% at $\alpha=50{,}000$. However, the additional
improvement became limited beyond $\alpha=5{,}000$, indicating
diminishing gains from further increasing the cyber-physical-risk scaling
factor.

\subsection{Sherlock Dataset Results}

As shown in Table~\ref{tab:alpha_improvement},
the maximum cost improvement increased from 0.00076\% at $\alpha=1$ to
0.07376\% at $\alpha=100$. No cost improvement was observed for
$\alpha\geq1{,}000$ across the evaluated FP/FN configurations.

\subsection{Discussion}

The experiments show that the effect of cyber-physical-risk-aware iterative
feature reweighting on operational cost depends on the underlying
dataset. For the ICS dataset, increasing $\alpha$ continued to improve
the minimum observed operational cost, although the incremental gains
became smaller at higher values. In contrast, the Sherlock dataset
showed increasing improvement up to $\alpha=100$, followed by no
observed improvement for $\alpha\geq1{,}000$.
These differences indicate that the operational effect of cyber-risk
scaling is dataset-dependent and may be influenced by the distribution
of misclassification risk across generators and its interaction with
the dispatch constraints. The two benchmarks differ in system
structure, feature representation, and generator-level sample
assignment, which may contribute to the different sensitivity patterns
observed in the experiments.
Overall, the results demonstrate that iterative feature reweighting can
produce lower operational dispatch costs than the corresponding
baseline for selected parameter configurations, while F1-score provides
a complementary measure of intrusion detection performance.

\section{Conclusion}

This paper presented an optimization-informed cyber-physical-risk framework that integrates machine learning-based intrusion detection with power grid operational optimization through iterative feature reweighting. The proposed approach transforms intrusion detection misclassification outcomes into generator-level cyber-physical-risk costs and incorporates these costs into the operational dispatch model. Previous logistic regression weights are then used as feature weights for subsequent intrusion detection iterations, enabling iterative interaction between intrusion detection and cyber-risk-aware operational optimization.
The proposed framework was evaluated using the Sherlock and Industrial Control System (ICS) power-system datasets under different cyber-physical-risk and misclassification-cost configurations. Experimental results showed that the effect of the framework on operational cost depends on the dataset and parameter configuration, with lower operational costs observed for selected configurations relative to their corresponding baselines. The hyperparameter sensitivity analysis further demonstrated different responses to the cyber-physical-risk scaling factor across the two datasets, while F1-score provided a complementary measure of intrusion detection performance.
Future work will investigate the proposed framework using larger power-system benchmarks, additional intrusion detection models, and additional operational constraints to further evaluate cyber-physical-risk-aware power grid operation.

\bibliographystyle{IEEEtran}
\bibliography{references}

@article{liu2011false,
  title={False data injection attacks against state estimation in electric power grids},
  author={Liu, Yao and Ning, Peng and Reiter, Michael K},
  journal={ACM Transactions on Information and System Security (TISSEC)},
  volume={14},
  number={1},
  pages={1--33},
  year={2011},
  publisher={ACM New York, NY, USA}
}

@article{krause2021cybersecurity,
  title={Cybersecurity in power grids: Challenges and opportunities},
  author={Krause, Tim and Ernst, Raphael and Klaer, Benedikt and Hacker, Immanuel and Henze, Martin},
  journal={Sensors},
  volume={21},
  number={18},
  pages={6225},
  year={2021},
  publisher={MDPI}
}

@article{sun2018cyber,
  title={Cyber security of a power grid: State-of-the-art},
  author={Sun, Chih-Che and Hahn, Adam and Liu, Chen-Ching},
  journal={International Journal of Electrical Power \& Energy Systems},
  volume={99},
  pages={45--56},
  year={2018},
  publisher={Elsevier}
}

@inproceedings{muller2022assessment,
  title={Assessment of cyber-physical intrusion detection and classification for industrial control systems},
  author={M{\"u}ller, Nils and Ziras, Charalampos and Heussen, Kai},
  booktitle={2022 IEEE International Conference on Communications, Control, and Computing Technologies for Smart Grids (SmartGridComm)},
  pages={432--438},
  year={2022},
  organization={IEEE}
}

@inproceedings{wagner2024sherlock,
  title={Sherlock: A Dataset for Process-aware Intrusion Detection Research on Power Grid Networks: Dataset Paper},
  author={Wagner, Eric and Bader, Lennart and Wolsing, Konrad and Serror, Martin},
  booktitle={Proceedings of the Fifteenth ACM Conference on Data and Application Security and Privacy},
  pages={419--424},
  year={2024}
}

@article{sahani2023machine,
  title={Machine learning-based intrusion detection for smart grid computing: A survey},
  author={Sahani, Nitasha and Zhu, Ruoxi and Cho, Jin-Hee and Liu, Chen-Ching},
  journal={ACM Transactions on Cyber-Physical Systems},
  volume={7},
  number={2},
  pages={1--31},
  year={2023},
  publisher={ACM New York, NY}
}

@ARTICLE{7063234,
  author={Pan, Shengyi and Morris, Thomas and Adhikari, Uttam},
  journal={IEEE Transactions on Smart Grid}, 
  title={Developing a Hybrid Intrusion Detection System Using Data Mining for Power Systems}, 
  year={2015},
  volume={6},
  number={6},
  pages={3104-3113},
  doi={10.1109/TSG.2015.2409775}
}

@article{haque2020machine,
  title={Machine learning in generation, detection, and mitigation of cyberattacks in smart grid: A survey},
  author={Haque, Nur Imtiazul and Shahriar, Md Hasan and Dastgir, Md Golam and Debnath, Anjan and Parvez, Imtiaz and Sarwat, Arif and Rahman, Mohammad Ashiqur},
  journal={arXiv preprint arXiv:2010.00661},
  year={2020}
}

@inproceedings{yang2020distributed,
  title={Distributed economic dispatch for cyber attacked smart grid based on resilient event-triggered consensus},
  author={Yang, Feisheng and Kang, Peipei and Guan, Xiaohong},
  booktitle={2020 59th IEEE Conference on Decision and Control (CDC)},
  pages={5725--5730},
  year={2020},
  organization={IEEE}
}

@article{sun2017robust,
  title={Robust optimization in electric power systems},
  author={Sun, XA and Terlaky, T and Anjos, MF and Ahmed, S},
  journal={Advances and Trends in Optimization with Engineering Applications},
  pages={357--365},
  year={2017},
  publisher={SIAM}
}

@article{tan2015integrity,
  title={Integrity attacks on real-time pricing in electric power grids},
  author={Tan, Rui and Krishna, Varun Badrinath and Yau, David KY and Kalbarczyk, Zbigniew},
  journal={ACM Transactions on Information and System Security (TISSEC)},
  volume={18},
  number={2},
  pages={1--33},
  year={2015},
  publisher={ACM New York, NY, USA}
}

@article{elmachtoub2022smart,
  title={Smart “predict, then optimize”},
  author={Elmachtoub, Adam N and Grigas, Paul},
  journal={Management Science},
  volume={68},
  number={1},
  pages={9--26},
  year={2022},
  publisher={INFORMS}
}

@article{mandi2024decision,
  title={Decision-focused learning: Foundations, state of the art, benchmark and future opportunities},
  author={Mandi, Jayanta and Kotary, James and Berden, Senne and Mulamba, Maxime and Bucarey, Victor and Guns, Tias and Fioretto, Ferdinando},
  journal={Journal of Artificial Intelligence Research},
  volume={80},
  pages={1623--1701},
  year={2024}
}

@article{donti2017task,
  title={Task-based end-to-end model learning in stochastic optimization},
  author={Donti, Priya and Amos, Brandon and Kolter, J Zico},
  journal={Advances in neural information processing systems},
  volume={30},
  year={2017}
}

@article{bertsimas2020predictive,
  title={From predictive to prescriptive analytics},
  author={Bertsimas, Dimitris and Kallus, Nathan},
  journal={Management Science},
  volume={66},
  number={3},
  pages={1025--1044},
  year={2020},
  publisher={INFORMS}
}

@inproceedings{elkan2001foundations,
  title={The foundations of cost-sensitive learning},
  author={Elkan, Charles},
  booktitle={International joint conference on artificial intelligence},
  volume={17},
  number={1},
  pages={973--978},
  year={2001},
  organization={Lawrence Erlbaum Associates Ltd}
}

@article{huang2020real,
  title={Real-time distributed economic dispatch scheme of grid-connected microgrid considering cyberattacks},
  author={Huang, Chongxin and Wang, Jing and Deng, Song and Yue, Dong},
  journal={IET Renewable Power Generation},
  volume={14},
  number={14},
  pages={2750--2758},
  year={2020},
  publisher={Wiley Online Library}
}

@article{li2025resilient,
  title={Resilient distributed economic dispatch for cyber--physical power systems considering carbon emissions trading and false data injection attacks},
  author={Li, Xue and Yang, Junlin and Du, Dajun and Zhou, Zhe and Li, Kang and Wu, Lei},
  journal={Energy},
  volume={332},
  pages={136881},
  year={2025},
  publisher={Elsevier}
}

@article{ding2020cyber,
  title={Cyber risks of PMU networks with observation errors: Assessment and mitigation},
  author={Ding, Weiyong and Xu, Maochao and Huang, Yu and Zhao, Peng},
  journal={Reliability Engineering \& System Safety},
  volume={198},
  pages={106873},
  year={2020},
  publisher={Elsevier}
}

@article{yan2012survey,
  title={A survey on cyber security for smart grid communications},
  author={Yan, Ye and Qian, Yi and Sharif, Hamid and Tipper, David},
  journal={IEEE communications surveys \& tutorials},
  volume={14},
  number={4},
  pages={998--1010},
  year={2012},
  publisher={IEEE}
}

@article{wlazlo2021man,
  title={Man-in-the-middle attacks and defence in a power system cyber-physical testbed},
  author={Wlazlo, Patrick and Sahu, Abhijeet and Mao, Zeyu and Huang, Hao and Goulart, Ana and Davis, Katherine and Zonouz, Saman},
  journal={IET Cyber-Physical Systems: Theory \& Applications},
  volume={6},
  number={3},
  pages={164--177},
  year={2021},
  publisher={Wiley Online Library}
}

@inproceedings{sourav2023machine,
  title={Machine learning assisted bad data detection for high-throughput substation communication},
  author={Sourav, Suman and Biswas, Partha P and Mohanraj, Vyshnavi and Chen, Binbin and Mashima, Daisuke},
  booktitle={ICC 2023-IEEE International Conference on Communications},
  pages={4131--4137},
  year={2023},
  organization={IEEE}
}

@misc{gurobi,
  author = {{Gurobi Optimization, LLC}},
  title = {{Gurobi Optimizer Reference Manual}},
  year = {2026},
  url = "https://www.gurobi.com"
}

\end{document}